\documentclass[aps,prl,twocolumn,groupedaddress,showpacs]{revtex4}
\usepackage[dvips]{graphicx}
\usepackage{natbib}
\newcommand{\vesc}{v_{\mathrm{esc}}}

\newcommand{\dd}{\mathrm{d}}

\newcommand{\rhox}{\rho_{\mathrm{x}}}

\newcommand{\nx}{n_{\mathrm{x}}}
\newcommand{\np}{n_{\mathrm{p}}}
\newcommand{\Nx}{N_{\mathrm{x}}}
\newcommand{\Np}{N_{\mathrm{p}}}

\newcommand{\Cx}{C_{\mathrm{x}}}

\newcommand{\Gammab}{\Gamma_{\mathrm{B}}}

\newcommand{\sigmap}{\sigma_{\mathrm{p}}}

\newcommand{\Mx}{M_{\mathrm{x}}}
\newcommand{\Tx}{T_{\mathrm{x}}}

\newcommand{\mn}{M_{\mathrm{N}}}

\newcommand{\mproton}{m_{\mathrm{p}}}

\newcommand{\kms}{\ ~\mathrm{km s}^{-1}}

\begin{document}

\title{
Asymmetric dark matter may alter the evolution of low-mass stars and brown dwarfs
}
\author{Andrew R. Zentner and Andrew P. Hearin}
\affiliation{
Department of Physics and Astronomy \& Pittsburgh Particle physics, Astrophysics, 
and Cosmology Center (PITT PACC), University of Pittsburgh, Pittsburgh, PA 15260, USA
}

\date{\today}

\begin{abstract}

We study energy transport by asymmetric dark matter in the interiors of 
very low-mass stars and brown dwarfs.  Our motivation is to explore astrophysical 
signatures of asymmetric dark matter, which otherwise may not be amenable to 
conventional indirect dark matter searches.  In viable models, the additional 
cooling of very low-mass stellar cores can alter stellar properties.  
Asymmetric dark matter with mass $4 \lesssim \Mx/\mathrm{GeV} \lesssim 10$ and a 
spin-dependent (spin-independent) cross section of 
$\sigma_{\mathrm{p}}^{\mathrm{SD}} \sim 10^{-37}\, \mathrm{cm}^2$ 
($\sigma_{\mathrm{p}}^{\mathrm{SI}} \sim 10^{-40}\, \mathrm{cm}^2$) 
can increase the minimum mass of main sequence hydrogen burning, partly determining 
whether or not the object is a star at all.  Similar dark matter candidates 
reduce the luminosities of low-mass stars and accelerate the cooling of brown dwarfs.  
Such light dark matter is of particular interest given results from the DAMA, CoGeNT, 
and CRESST dark matter searches.  We discuss possibilities for observing dark matter effects 
in stars in the solar neighborhood, globular clusters, and, of particular promise, local dwarf 
galaxies, among other environments, as well as exploiting these effects to constrain 
dark matter properties.

\end{abstract}


\maketitle

%%%%%%%%%%%%%%%%%%%%%%%%%%%%%%% intro

Overwhelming evidence indicates that a form of non-baryonic matter 
constitutes the majority of mass in the Universe.  
The unknown nature of the {\em dark matter} (DM) is a 
fundamental problem in cosmology and particle physics.  
Among DM candidates, weakly-interacting massive particles (WIMPs), 
particularly the lightest superpartners in supersymmetric theories, 
have garnered the most attention \cite{jungman_etal96}.  
We study the effects of asymmetric dark matter (ADM) 
on the evolution of very low-mass (VLM) stars and brown dwarfs (BD).  
We show that stars with masses $M_* \lesssim 0.15 M_{\odot}$ and BDs 
just below the minimum mass for a hydrogen-burning main sequence (MS) star, 
$M \gtrsim 0.05 M_{\odot}$, may have their evolution significantly altered 
by the accumulation of DM.

ADM has a relic asymmetry, so it does not 
annihilate as thermal relic WIMPs do.  
Consequently, ADM is not amenable to indirect 
detection via observations of annihilation 
products from astrophysical sources \cite{porter_etal11}.  
Our result suggests a future indirect identification 
method for ADM in astrophysical sources.  
In the present context, the relic asymmetry allows stars to collect large 
amounts of DM without the accumulation being moderated by 
annihilation \cite{griest_seckel87,taoso_etal10,frandsen_sarkar10}.  
ADM models offer a possible explanation 
for the DM and baryon densities of the Universe being of the same order 
\cite{barr_etal90,kaplan92,hooper_etal05,gudnason_etal06,roszkowski_seto07,
kaplan_etal09,sannino_zwicky09,foadi_etal09,cohen_etal10,davoudiasl_etal10,
falkowski_etal11}.  ADM models are relevant to DM particles with 
masses $\Mx \lesssim 15~\mathrm{GeV}$, 
lower than typical WIMPs, and interest in 
ADM has been fueled by possible direct detection signals indicating 
low-mass DM by the DAMA \cite{dama08}, CoGeNT \cite{cogent11,cogent11b} and 
CRESST-II \cite{cresst11} experiments \cite{fitzpatrick_etal10,chang_etal10} 
(though challenged by Xenon-10 limits, Ref.~\cite{xenon10_11}).

It has long been recognized that DM could accumulate 
in the Sun (and other stars) subtly altering its properties, 
particularly solar neutrino fluxes 
\cite{spergel_press85,gould92,cumberbatch_etal10,taoso_etal10,frandsen_sarkar10,mcdermott_etal11}.  
The scenario is simple.  As the star orbits in the DM halo, 
some DM particles scatter off stellar nuclei and become bound to the star.  
The captured DM particles can be non-negligible contributors 
to energy transport in the stellar interior.

DM with $\Mx \lesssim 15\,\mathrm{GeV}$ is captured by stars at a rate 
\begin{eqnarray}
\label{eq:caprate}
\Cx & \approx & C_{\odot}^{\mathrm{SI},\mathrm{SD}}\, 
\left(\frac{\rhox}{0.4\,\mathrm{GeV}\,\mathrm{cm}^{-3}}\right)\,
\left(\frac{\sigmap}{10^{-43}\, \mathrm{cm}^2} \right)\, \nonumber \\
    & \times & \left(\frac{\vesc}{618 \kms} \right)^2\,
\left(\frac{270 \kms}{\bar{v}}\right)
\left(\frac{M_*}{M_{\odot}} \right)\ ,
\end{eqnarray}
where $M_*$ is stellar mass, 
$\rhox$ is the DM density in the star's vicinity, 
$\sigmap$ is the cross-section for ADM-proton scattering, 
$\vesc$ is the stellar escape speed, 
and $\bar{v}$ is the typical speed at infinity of infalling DM 
particles.  The coefficients 
$C_{\odot}^{\mathrm{SI}} \approx 7 \times 10^{22}\, \mathrm{s}^{-1}$ and 
$C_{\odot}^{\mathrm{SD}} \approx 5 \times 10^{21}\, (5\, \mathrm{GeV}/\Mx)\, \mathrm{s}^{-1}$ 
give the rates in the Sun due to spin-independent (SI) and spin-dependent (SD) 
scattering respectively \cite{gould92}.  We include the $\Mx$ dependence in 
$C_{\odot}^{\mathrm{SD}}$.  $C_{\odot}^{\mathrm{SI}}$ is a weak function of $\Mx$. 
In the SI case, we take the scattering cross section on a nucleus 
with mass $\mn$ and mass number $A$ to be 
$\sigma_{\mathrm{N}} = \sigmap^{\mathrm{SI}}\, A^2\, \mn^2(\Mx+\mproton)^2/(\Mx+\mn)^2\mproton^2$, 
where $\mproton$ is the proton mass.  
We provide Eq.~(\ref{eq:caprate}) for convenience, but calculate time-dependent 
capture rates using the full formulae from \cite{gould92} as described 
in Ref.~\cite{zentner09}.

VLM stars and BDs are interesting DM laboratories 
for several reasons.  First, dark matter capture 
rates depend upon stellar structure only insomuch as 
$\vesc$ does.  In VLM stars, mass is proportional to radius 
so VLM stars have similar escape speeds to the Sun and 
capture nearly the same amount of DM per unit mass.  
Second, stellar luminosity scales with mass roughly as 
$L \propto M_*^{3}$ for VLM stars, so a $M_* \sim 0.1M_{\odot}$ 
star radiates only $L \sim 10^{-3}L_{\odot}$, and DM needs to transport 
a relatively small energy flux to alter the stellar evolution.  
Third, VLM stars have core temperatures many times lower 
than the Sun and nuclear burning rates in this regime 
are rapidly-varying functions of temperature, so small 
temperature changes have large effects on luminosity.  
These considerations suggest that DM energy 
transport may dramatically alter VLM stars.

The mean free path of captured DM exceeds stellar radii, 
so DM energy transport is {\em non local}.  
Nevertheless, we can make an order-of-magnitude estimate 
of the energy flux transmitted by DM from the stellar core 
by considering this transport to be 
diffusive, with an effective diffusion coefficient  
$\eta' \sim \eta (\sigmap/\sigma_{\mathrm{c}})^2$, where 
$\eta \sim (1/\np \sigmap) \sqrt{kT_{\mathrm{c}}/\Mx}$ is the 
standard for diffusive transport \cite{press_spergel85}.  
The cross-section for which the star is 
optically thick is $\sigma_{\mathrm{c}} \equiv (\mproton/M_*) \pi R_*^2$.  
The factors $(\sigmap/\sigma_{\mathrm{c}})^2$ account for the 
facts that DM orbits span far less than a mean 
free path and DM particles scatter only after many orbits.  
Using $\eta'$ and estimating the temperature gradient as 
$\dd T/\dd r \sim T_{\mathrm{c}}/R_*$, the rate at which 
DM removes energy from the stellar core is, 
\begin{equation}
\label{eq:lsimple}
\frac{L_{\mathrm{x}}}{10^{-3}L_{\odot}} \sim 1.6 \times 10^{13}\, 
\left(\frac{\sigmap^{\mathrm{SD}}}{10^{-37}\, \mathrm{cm}^2}\right) \left( \frac{\Nx}{\Np} \right) 
\sqrt{\frac{\mproton}{\Mx}}.
\end{equation}
Eq.~(\ref{eq:lsimple}) includes only SD scattering for simplicity, 
$\Nx$ is the number of captured DM particles, and $\Np$ is the number of protons.   
A luminosity of $L_* \sim 10^{-3} L_{\odot}$ is typical for the 
$M_* \approx 0.1 M_{\odot}$ star we have assumed.

The energy flux carried by DM is of order the 
stellar luminosity ($\sim 10^{-3}L_{\odot}$) if 
$\Nx \sim 10^{-13} \Np \sim 10^{43}$.  
A cross section of $\sigma_{\mathrm{p}}^{\mathrm{SD}}=10^{-37}\, \mathrm{cm}^2$ yields a 
capture rate of $\Cx \sim 5 \times 10^{26}\, \mathrm{s}^{-1}$.  Assuming a lifetime 
$\tau \sim 10^{10}\, \mathrm{yr}$, then $\Nx \sim 10^{44}$, so large effects are 
possible in ADM models.  We have computed the DM heat transfer rate 
within $n=3/2$ polytropic stellar models, appropriate for 
fully convective stars with $M_* \sim 0.1 M_{\odot}$ \cite{kippenhahn_weigert}, and 
confirmed this approximate result.  These approximations are not self-consistent; 
however, they provide strong indications that ADM effects may be non-negligible.
DM effects are dramatic in environments with large capture rates. 
We parametrize environment with $\Gammab$, the ratio of the capture rate of DM 
to the {\em standard capture rate in the solar neighborhood}.  
As we discuss below, environments with $\Gammab \gg 1$ are known to 
exist and $\Gammab \gg 1$ may effectively be realized in the solar neighborhood.

We perform a self-consistent stellar evolution calculation 
by computing capture rates as in Ref.~\cite{zentner09} and including 
DM heat transport in the Modules for Experiments in
Stellar Astrophysics (MESA) software \cite{mesa_instrument}.  
We base our models on the MESA VLM star models (\S~7.1 of Ref.~\cite{mesa_instrument}).  
Ideally, one would calculate DM energy transport by solving the 
Boltzmann equation at each stage of stellar evolution 
\cite{gould_raffelt90b}, 
but this is computationally challenging and beyond our 
present scope.  We adopt the approximations of Ref.~\cite{spergel_press85}, 
namely, that DM has an effective temperature $\Tx$, and 
that energy transport can be estimated from the first moment of 
the Boltzmann equation.  In the case of {\em SD} 
scattering off of hydrogen, the energy per unit mass transmitted from 
baryons to DM is 
\begin{equation}
\label{eq:eps}
\epsilon = 8\sqrt{\frac{2}{\pi}}\, \frac{\nx \sigma_{\mathrm{p}}^{\mathrm{SD}} \Mx}{(\Mx + \mproton)^2}\, 
\frac{f_{\mathrm{H}}}{\mproton}\, 
\left( \frac{\mproton \Tx + \Mx T}{\Mx \mproton} \right)^2\, (T-\Tx),
\end{equation}
where $\nx$ is the local DM number density, 
$f_{\mathrm{H}}=0.71$ is the local mass fraction in hydrogen, 
and $\Tx$ is fixed by requiring that there be no {\em net} energy transfer.  
In the case of {\em SI} scattering, one sums Eq.~(\ref{eq:eps}) over all nuclei. 
Eq.~\ref{eq:eps} has known shortcomings, 
but corrections are generally of order unity 
and there is no general treatment that 
can be incorporated into a 
stellar evolution model in a 
computationally-feasible manner \cite{gould_raffelt90b}.  
Uncertainty associated with DM parameters is large, 
so this approximation should suffice for the present purpose.

%----------------
% example of central density and temperature
%----------------
\begin{figure}[t!]
\includegraphics[height=8.5cm]{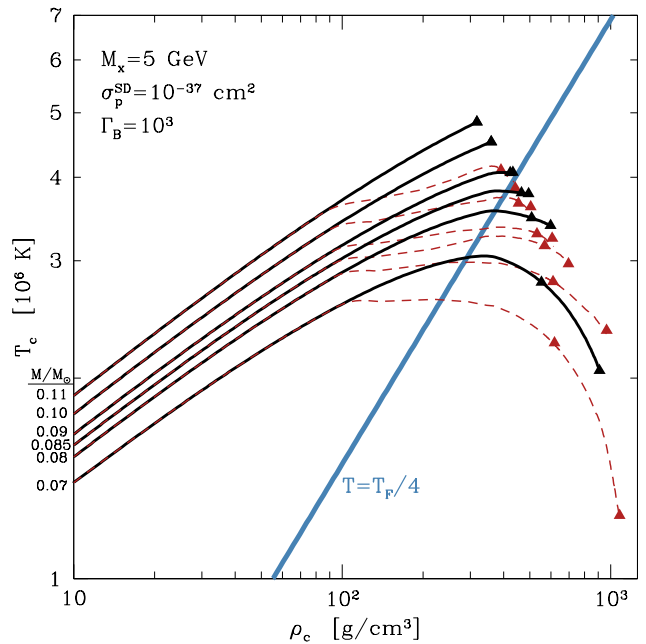}
%\vspace*{-16pt}
\caption{
The evolution of central temperature $T_{\mathrm{c}}$ as a function of 
central density $\rho_{\mathrm{c}}$ for VLM stars with masses 
labeled at left. {\em Solid} lines are standard stellar models.  
{\em Dashed} lines include dark matter cooling.  The line at which 
temperature is four times the Fermi temperature is marked, below which 
electron degeneracy pressure dominates pressure support.  
Triangles mark ages of $10^9$~yr and $5 \times 10^9$~yr.  
Trajectories with one triangle exhibit no evolution during this period and 
are on the stellar MS.
}
\label{fig:rhoctc}
\vspace*{-14pt}
\end{figure}
%---------------------------------

In the interest of brevity, we quote detailed results for a fiducial model of 
ADM with $\Mx=5\, \mathrm{GeV}$ and 
$\sigma_{\mathrm{p}}^{\mathrm{SD}}=10^{-37}\, \mathrm{cm}^2$, 
well below current limits \cite{picasso09}.  However, we obtain similar results for 
SI cross-sections $\sigma_{\mathrm{p}}^{\mathrm{SI}} \sim 10^{-40}\, \mathrm{cm}^2$ 
due to the large cross sections for DM scattering on heavier elements, 
primarily He, C, O, Ne.  Specifically, a model with $\Mx \approx 7\, \mathrm{GeV}$ and 
$\sigma_{\mathrm{p}}^{\mathrm{SI}} \sim 10^{-40}\, \mathrm{cm}^2$ 
yields stellar evolution similar to our fiducial case 
and is of interest given indications of low-mass DM scattering by the 
DAMA \cite{dama08} and CoGeNT \cite{cogent11,cogent11b} 
collaborations.  We account for possible loss of DM by evaporation from 
the stellar interior following Ref.~\cite{gould87a}, with the 
result that evaporation is negligible for DM masses $\Mx \gtrsim 3\, \mathrm{GeV}$.  
This is slightly smaller than the evaporation mass from the Sun 
($\Mx \gtrsim 3.7\, \mathrm{GeV}$), primarily due to the cooler interiors of VLM stars.

Fig.~\ref{fig:rhoctc} shows the evolution of central density $\rho_{\rm c}$, 
and central temperature $T_{\rm c}$, for VLM stars and BDs.  
This classic plot illustrates that a collapsing gas cloud of 
sufficiently low mass will achieve a maximum $T_{\rm c}$ too meager 
to ignite hydrogen burning at a level that can halt gravitational collapse \cite{kumar63}.  
Maximum $T_{\rm c}$ is achieved when pressure support becomes 
dominated by electron degeneracy.  Objects with $M_* \gtrsim 0.08M_{\odot}$ 
halt contraction when hydrogen burning begins and 
enter the stellar main sequence, enjoying long MS lifetimes. 
Lower-mass objects continue to contract and cool, becoming BDs.  
In Fig.~\ref{fig:rhoctc}, we compare standard stellar evolution to evolution 
including DM cooling in our fiducial ADM model with a capture 
rate boosted by $\Gammab=10^3$.  In the DM models, the additional cooling 
causes {\em degeneracy to be achieved at lower densities} and 
temperature maxima are reduced at fixed stellar mass.  
Consequently, the minimum mass for MS H-burning increases by $\sim 15\%$.

We emphasize, with regard to Fig.~\ref{fig:rhoctc}, that the 
primary effect of ADM cooling is to drive the stellar core to degeneracy 
at a lower density.  Once degeneracy sets in, DM energy 
transport becomes significantly less important and the degenerate core 
cools and contracts as in the standard evolution of a 
degenerate object \cite{kippenhahn_weigert,chabrier_baraffe00}.  At all 
stages of evolution the stellar temperature profile is monotonic and the 
DM temperature closely tracks the baryonic temperature of the stellar core.  
Moreover, convective energy transport (see \cite{kippenhahn_weigert,chabrier_baraffe00}) 
remains an important channel for energy transport in the bulk of the stellar interiors 
in all of the models we have considered.

\begin{figure}[t!]
\includegraphics[height=8.5cm]{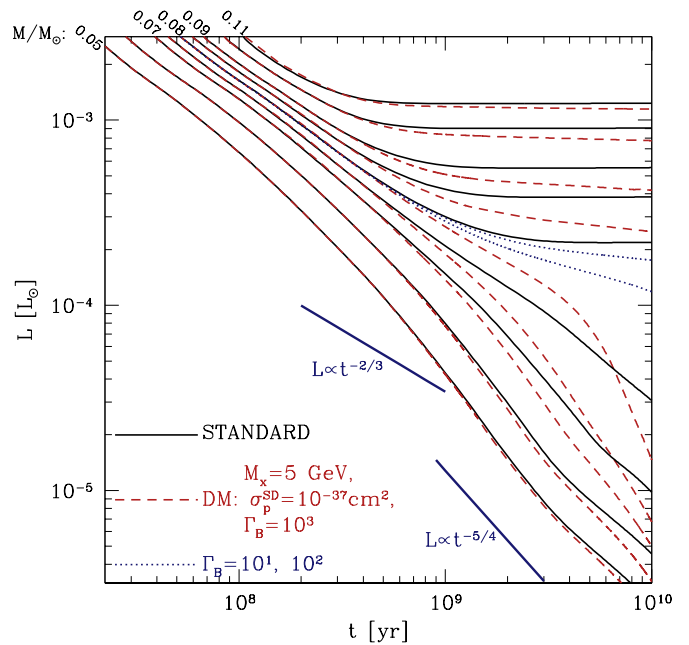}
\caption{
Luminosity evolution of low-mass stars and brown dwarfs.  
{\em Solid} and {\em Dashed} are as in Fig.~\ref{fig:rhoctc}.  
{\em Dotted} lines show evolution for 
$\Gammab = 10^2,\,10^1$ for the $M=0.08M_{\odot}$ models.  
The masses of the objects are 
$M=0.11M_{\odot}$, $0.1M_{\odot}$, $0.09M_{\odot}$, $0.085M_{\odot}$, $0.08M_{\odot}$, $0.075M_{\odot}$, $0.07M_{\odot}$, 
$0.06M_{\odot}$, $0.05M_{\odot}$ from top to bottom 
(every other line is labeled at the upper left).  
We show power laws for cooling of a fully convective object 
at constant effective temperature ($L \propto t^{-2/3}$) and a fully degenerate 
object ($L \propto t^{-5/4}$) for reference (e.g., Ref.~\cite{chabrier_baraffe00}).
}
\label{fig:lumevol}
\vspace*{-14pt}
\end{figure}

Important consequences of DM cooling are more apparent in 
Fig.~\ref{fig:lumevol}, which depicts the evolution of luminosity 
for collapsing objects with $0.06 \le M_*/M_{\odot} \le 0.11M_{\odot}$.  
In standard models, objects with $M_* \gtrsim 0.08M_{\odot}$ enter a 
long-lived phase of constant luminosity supported by core H-burning, the MS.  
Lower-mass objects cool and dim incessantly.  In our DM model, 
objects with $M_* \lesssim 0.1M_{\odot}$ dim continually 
and exhibit {\em no constant-luminosity phase}, so 
objects with $0.08 \lesssim M/M_{\odot} \lesssim 0.10$ 
{\em are not MS stars}.  DM cooling effects are most dramatic just 
below the minimum MS mass, $M_* \approx 0.08M_{\odot}$.   
In the standard case, such objects have their dimming delayed 
by non-negligible nuclear burning, yet nuclear reaction rates at 
such low temperatures are sensitive functions of temperature 
\cite{kippenhahn_weigert}, so DM cooling quells this H-burning and drives 
these objects to lower luminosities.  
At $M=0.08M_{\odot}$ the object with DM cooling is a factor of $\sim 15$ less 
luminous than its standard counterpart after $10^{10}\, \mathrm{yr}$.  Notice that 
DM has little effect on objects with $M \lesssim 0.05M_{\odot}$ because these objects 
are {\em never} significantly affected by nuclear burning.  
Fig.~\ref{fig:lumevol} shows two additional cases of an 
$M=0.08M_{\odot}$ object with $\Gammab=10$ and $\Gammab=10^2$, illustrating that 
DM cooling can have appreciable effects with more modest DM capture rates.

%----------------
% example of luminosity evolution
%----------------
%---------------------------------

\begin{figure}[t!]
\includegraphics[height=8.4cm]{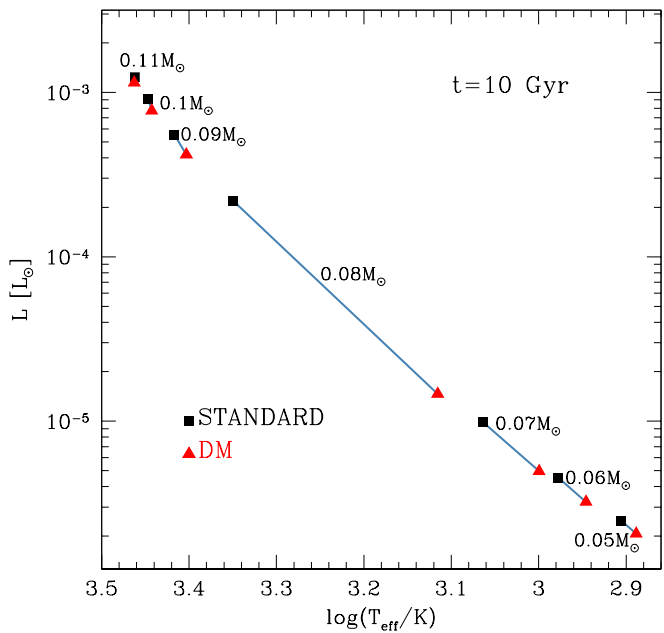}
\caption{
Shifts in stellar positions on an H-R diagram for our fiducial 
dark matter model.  Squares are standard evolution while triangles 
are the fiducial ADM model.  We show shifts evaluated at six masses 
to avoid clutter in the diagram.  Lines connect the standard and 
ADM cases at each mass.
}
\label{fig:hr}
\vspace*{-14pt}
\end{figure}

Figure~\ref{fig:hr} displays the shifts in the positions of 
stars on the HR diagram at several values of stellar mass after 
10~Gyr of stellar evolution.  The changes in effective temperature 
and luminosity are significant for a range of stellar masses 
from $0.05-0.10\,\mathrm{M}_{\odot}$.  ADM 
accelerates the cooling and dimming of these stars, giving rise 
to a dearth of relatively luminous VLM stars and BDs.

\begin{figure}[t!]
\includegraphics[height=8.00cm]{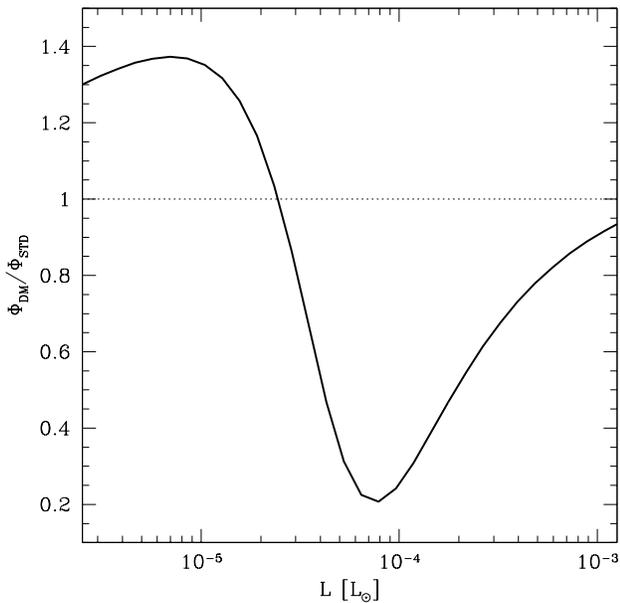}
\caption{
Relative stellar/BD luminosity function for a 
stellar population at an age of 10~Gyr for our 
fiducial ADM model, $\Phi_{\rm ADM}(L)$, compared 
to the luminosity function in the case of standard 
stellar evolution, $\Phi_{\rm STD}(L)$, assuming 
an identical IMF.  
}
\label{fig:lf}
\vspace*{-14pt}
\end{figure}

It is possible to approximate the deficit of VLM stars.  
Assuming a fixed initial mass function and stellar metallicity, 
the change in the luminosity function of VLM stars and BDs is determined by the 
mass-luminosity relation, $L(M)$.  We have approximated $L(M)$ using 
stellar models computed for masses separed by $\Delta M = 0.005\,\mathrm{M}_{\odot}$ 
between $M=0.04\,\mathrm{M}_{\odot}$ and $M=0.12\,\mathrm{M}_{\odot}$.  
Figure~\ref{fig:lf} shows the ratio of the luminosity function in 
our fiducial ADM scenario to a standard luminosity function for a 
stellar population after 10~Gyr of evolution.  The significant 
deficit of VLM stars and BDs with luminosities 
$2\times 10^{-5}\,\mathrm{L}_{\odot} \lesssim L \lesssim 10^{-3}\,\mathrm{L}_{\odot}$ 
is apparent.  Furthermore, an excess number of dim BDs near 
$L \sim 10^{-5}\,\mathrm{L}_{\odot}$ is also evident.  This 
pile-up results from the accelerated dimming and cooling low-mass 
objects in the ADM model that would otherwise be more luminous 
VLM stars and BDs.

%----------------
% example of HR diagram
%----------------
%---------------------------------

The effects of DM on VLM stars and BDs may have interesting consequences.  
To be sure, observing VLM stars and BDs is challenging; however, 
several operating and forthcoming astronomical facilities count 
observations of local {\em and distant} VLM stars and BDs among their 
science drivers including 
PanSTARRS \cite{panstarrs02}, 
LSST \cite{lsstbook09}, 
Euclid \cite{euclid10}, 
TMT \footnote{{\tt http://www.tmt.org/}}, 
GMT \footnote{{\tt http://www.gmto.org/}}, 
and JWST \cite{jwstbd09}.  
At minimum, we have demonstrated that DM may have non-negligible effects on VLM stars and BDs, 
rendering them cooler and dimmer than otherwise at fixed age and chemical composition.  
The luminosity function of VLM stars and BDs should become significantly 
shallower from $10^{-3} \gtrsim L/L_{\odot} \gtrsim 10^{-5}$ and steepen at lower luminosities as 
indicated by Fig.~\ref{fig:lumevol} and Fig.~\ref{fig:lf}.  It may also be possible to 
identify seemingly anomalous VLM stellar or BD companions in studies of transits of 
more luminous stars.  The strength of this effect should be 
{\em correlated with environment}: it is dramatic in regions of high-DM density and small 
in regions of low-DM density.  This correlation may distinguish DM influences on stellar 
evolution from uncertainties in stellar modeling.

Dramatic effects generally require capture rates larger than {\em standard} rates 
in the solar neighborhood.  High capture rates are possible in the solar 
neighborhood in particular models.  
A co-rotating disk of DM near the Galactic plane is a 
prediction of hierarchical galaxy formation and may provide a boost 
of $\Gammab \sim 10$ \cite{bruch_etal09,purcell_etal09}.  Alternatively, DM with a 
significant self-interaction cross section can enhance capture rates equivalent to 
$\Gammab \sim 10^2$ or greater in the Sun and nearby stars \cite{zentner09,frandsen_sarkar10}.  
This suggests that observations of local VLM stars and BDs may constrain 
models of self-interacting ADM, though such constraints require careful 
modeling of stellar atmospheres and an exploration of stellar parameters, such as metallicity.  
Additionally, trace populations of VLM stars that were either (1) 
liberated in the disruption of merging dwarf galaxies during hierarchical formation of 
the Milky Way or (2) members of the halo star population that orbit through 
the Galactic Center would still be effected by ADM accumulated while in these 
higher-density environments and be subject to the DM effects we describe.

Globular clusters (GC) are interesting environments in which to seek DM effects.  The 
formation, evolution, and DM content of GCs are still under debate.  
However, GCs are many orders of magnitude denser 
than the solar neighborhood and have internal velocity dispersions a factor of $\gtrsim 30$ 
less than the local value \cite{meylan_heggie97}.  Moreover, the bounds on their DM contents are weak, 
with the ratio of DM-to-stellar mass $M_{\mathrm{DM}}/M_{\mathrm{s}} \lesssim 1$ \cite{conroy_etal10}.  
Consequently, values of $M_{\mathrm{DM}}/M_{\mathrm{s}}$ well below those within reach of 
stellar kinematical studies would lead to values of $\Gammab \propto \rhox/\bar{v} \gg 1$ 
[see Eq.~(\ref{eq:caprate})].  This suggests that DM-induced alterations 
to stellar evolution may occur in GCs.  If consensus ever formed 
around evidence for low-mass ADM, observations of stellar populations 
in GCs may reveal their past or present DM contents.

For fixed DM parameters, high capture rates {\em should be realized in particular 
environments}.  One such environment is a Milky Way dwarf satellite galaxy.  
Consider the closest dwarf satellite, Segue I.  Stellar kinematics indicate 
that the DM density in Segue I is $\sim 10-10^2$ times the local value, 
while typical velocities are $10^2$ times lower \cite{simon_etal11}, 
suggesting $\Gammab \propto \rhox/\bar{v} \sim 10^3-10^4$.  VLM stars in 
dwarf galaxies are senstive to low-mass ADM and may be within reach of future 
observatories such as JWST.

%----------------
% example of Luminosity Function
%----------------
%---------------------------------

Interestingly, it may be possible to identify DM effects in the spectra of specific 
galaxies or GCs.  Ref.~\cite{vandokkum_conroy10} emphasized that 
spectral features due to Na and FeH are produced only by VLM stars and that the prevalence of 
these features reveals the contribution of VLM stars to the total galactic light.  
This suggests that stellar population synthesis studies may bear on the identification of DM and vice versa.  
Particular objects with kinematical or other evidence for high $\Gammab$ could 
be targeted for observation and the relative light contributed by VLM stars could 
limit the influence of DM on VLM stellar evolution.  Alternatively, though more speculatively, 
it is thought that the first generations of stars formed in the inner regions of early-forming 
dark matter halos, environments with significantly higher DM density than should accompany 
contemporary star formation in the Milky Way, so galaxies 
in which a significant amount of star formation occured at high redshift may have had 
low-mass stellar populations markedly altered by DM.

Several advances must be made before DM constraints based on this effect could be realized.  
First and foremost, observations of VLM stars and BDs must be improved.  Theoretically, 
it is necessary to implement computationally-efficient models of 
DM energy transport, to account for uncertainties in VLM stellar atmospheres, 
and explore stellar parameters, such as metallicity.  We will 
report on these efforts in a detailed follow-up study.  
That ADM may have potentially-observable 
effects on VLM stars and BDs opens up intriguing new avenues for 
constraining DM and learning about the environments of stellar populations.

%%%%%%%%%%%%%%%%%%%%%%%%%%%%%%%%%%%%%%%%%%

\begin{acknowledgments}

We thank Dan Boyanovsky, John Hillier, Aravind Natarajan, 
and Jeff Newman for helpful discussions.  We are grateful to 
B. Paxton, L. Bildsten, A. Dotter, F. Herwig, P. Lesaffre, and F. Timmes 
for making MESA publicly available. This work was supported by 
National Science Foundation grant PHY 0968888.

\end{acknowledgments}

\bibliography{ms}

\begin{thebibliography}{46}
\expandafter\ifx\csname natexlab\endcsname\relax\def\natexlab#1{#1}\fi
\expandafter\ifx\csname bibnamefont\endcsname\relax
  \def\bibnamefont#1{#1}\fi
\expandafter\ifx\csname bibfnamefont\endcsname\relax
  \def\bibfnamefont#1{#1}\fi
\expandafter\ifx\csname citenamefont\endcsname\relax
  \def\citenamefont#1{#1}\fi
\expandafter\ifx\csname url\endcsname\relax
  \def\url#1{\texttt{#1}}\fi
\expandafter\ifx\csname urlprefix\endcsname\relax\def\urlprefix{URL }\fi
\providecommand{\bibinfo}[2]{#2}
\providecommand{\eprint}[2][]{\url{#2}}

\bibitem[{\citenamefont{{Jungman} et~al.}(1996)\citenamefont{{Jungman},
  {Kamionkowski}, and {Griest}}}]{jungman_etal96}
\bibinfo{author}{\bibfnamefont{G.}~\bibnamefont{{Jungman}}},
  \bibinfo{author}{\bibfnamefont{M.}~\bibnamefont{{Kamionkowski}}},
  \bibnamefont{and} \bibinfo{author}{\bibfnamefont{K.}~\bibnamefont{{Griest}}},
  \bibinfo{journal}{\physrep} \textbf{\bibinfo{volume}{267}},
  \bibinfo{pages}{195} (\bibinfo{year}{1996}).

\bibitem[{\citenamefont{{Porter} et~al.}(2011)\citenamefont{{Porter},
  {Johnson}, and {Graham}}}]{porter_etal11}
\bibinfo{author}{\bibfnamefont{T.~A.} \bibnamefont{{Porter}}},
  \bibinfo{author}{\bibfnamefont{R.~P.} \bibnamefont{{Johnson}}},
  \bibnamefont{and} \bibinfo{author}{\bibfnamefont{P.~W.}
  \bibnamefont{{Graham}}} (\bibinfo{year}{2011}), \eprint{[arXiv:1104.2836]}.

\bibitem[{\citenamefont{{Griest} and {Seckel}}(1987)}]{griest_seckel87}
\bibinfo{author}{\bibfnamefont{K.}~\bibnamefont{{Griest}}} \bibnamefont{and}
  \bibinfo{author}{\bibfnamefont{D.}~\bibnamefont{{Seckel}}},
  \bibinfo{journal}{Nuclear Physics B} \textbf{\bibinfo{volume}{283}},
  \bibinfo{pages}{681} (\bibinfo{year}{1987}).

\bibitem[{\citenamefont{{Taoso} et~al.}(2010)}]{taoso_etal10}
\bibinfo{author}{\bibfnamefont{M.}~\bibnamefont{{Taoso}}} \bibnamefont{et~al.},
  \bibinfo{journal}{\prd} \textbf{\bibinfo{volume}{82}},
  \bibinfo{pages}{083509} (\bibinfo{year}{2010}).

\bibitem[{\citenamefont{{Frandsen} and {Sarkar}}(2010)}]{frandsen_sarkar10}
\bibinfo{author}{\bibfnamefont{M.~T.} \bibnamefont{{Frandsen}}}
  \bibnamefont{and} \bibinfo{author}{\bibfnamefont{S.}~\bibnamefont{{Sarkar}}},
  \bibinfo{journal}{Phys. Rev. Lett.} \textbf{\bibinfo{volume}{105}},
  \bibinfo{pages}{011301} (\bibinfo{year}{2010}).

\bibitem[{\citenamefont{{Barr} et~al.}(1990)\citenamefont{{Barr}, {Sekhar
  Chivukula}, and {Farhi}}}]{barr_etal90}
\bibinfo{author}{\bibfnamefont{S.~M.} \bibnamefont{{Barr}}},
  \bibinfo{author}{\bibfnamefont{R.}~\bibnamefont{{Sekhar Chivukula}}},
  \bibnamefont{and} \bibinfo{author}{\bibfnamefont{E.}~\bibnamefont{{Farhi}}},
  \bibinfo{journal}{Phys. Lett. B} \textbf{\bibinfo{volume}{241}},
  \bibinfo{pages}{387} (\bibinfo{year}{1990}).

\bibitem[{\citenamefont{{Kaplan}}(1992)}]{kaplan92}
\bibinfo{author}{\bibfnamefont{D.~B.} \bibnamefont{{Kaplan}}},
  \bibinfo{journal}{Phys. Rev. Lett.} \textbf{\bibinfo{volume}{68}},
  \bibinfo{pages}{741} (\bibinfo{year}{1992}).

\bibitem[{\citenamefont{{Hooper} et~al.}(2005)\citenamefont{{Hooper},
  {March-Russell}, and {West}}}]{hooper_etal05}
\bibinfo{author}{\bibfnamefont{D.}~\bibnamefont{{Hooper}}},
  \bibinfo{author}{\bibfnamefont{J.}~\bibnamefont{{March-Russell}}},
  \bibnamefont{and} \bibinfo{author}{\bibfnamefont{S.~M.}
  \bibnamefont{{West}}}, \bibinfo{journal}{Phys. Lett. B}
  \textbf{\bibinfo{volume}{605}}, \bibinfo{pages}{228} (\bibinfo{year}{2005}).

\bibitem[{\citenamefont{{Gudnason} et~al.}(2006)\citenamefont{{Gudnason},
  {Kouvaris}, and {Sannino}}}]{gudnason_etal06}
\bibinfo{author}{\bibfnamefont{S.~B.} \bibnamefont{{Gudnason}}},
  \bibinfo{author}{\bibfnamefont{C.}~\bibnamefont{{Kouvaris}}},
  \bibnamefont{and}
  \bibinfo{author}{\bibfnamefont{F.}~\bibnamefont{{Sannino}}},
  \bibinfo{journal}{\prd} \textbf{\bibinfo{volume}{73}},
  \bibinfo{pages}{115003} (\bibinfo{year}{2006}).

\bibitem[{\citenamefont{{Roszkowski} and {Seto}}(2007)}]{roszkowski_seto07}
\bibinfo{author}{\bibfnamefont{L.}~\bibnamefont{{Roszkowski}}}
  \bibnamefont{and} \bibinfo{author}{\bibfnamefont{O.}~\bibnamefont{{Seto}}},
  \bibinfo{journal}{Phys. Rev. Lett.} \textbf{\bibinfo{volume}{98}},
  \bibinfo{pages}{161304} (\bibinfo{year}{2007}).

\bibitem[{\citenamefont{{Kaplan} et~al.}(2009)\citenamefont{{Kaplan}, {Luty},
  and {Zurek}}}]{kaplan_etal09}
\bibinfo{author}{\bibfnamefont{D.~E.} \bibnamefont{{Kaplan}}},
  \bibinfo{author}{\bibfnamefont{M.~A.} \bibnamefont{{Luty}}},
  \bibnamefont{and} \bibinfo{author}{\bibfnamefont{K.~M.}
  \bibnamefont{{Zurek}}}, \bibinfo{journal}{\prd}
  \textbf{\bibinfo{volume}{79}}, \bibinfo{pages}{115016}
  (\bibinfo{year}{2009}).

\bibitem[{\citenamefont{{Sannino} and {Zwicky}}(2009)}]{sannino_zwicky09}
\bibinfo{author}{\bibfnamefont{F.}~\bibnamefont{{Sannino}}} \bibnamefont{and}
  \bibinfo{author}{\bibfnamefont{R.}~\bibnamefont{{Zwicky}}},
  \bibinfo{journal}{\prd} \textbf{\bibinfo{volume}{79}},
  \bibinfo{pages}{015016} (\bibinfo{year}{2009}).

\bibitem[{\citenamefont{{Foadi} et~al.}(2009)\citenamefont{{Foadi}, {Frandsen},
  and {Sannino}}}]{foadi_etal09}
\bibinfo{author}{\bibfnamefont{R.}~\bibnamefont{{Foadi}}},
  \bibinfo{author}{\bibfnamefont{M.~T.} \bibnamefont{{Frandsen}}},
  \bibnamefont{and}
  \bibinfo{author}{\bibfnamefont{F.}~\bibnamefont{{Sannino}}},
  \bibinfo{journal}{\prd} \textbf{\bibinfo{volume}{80}},
  \bibinfo{pages}{037702} (\bibinfo{year}{2009}).

\bibitem[{\citenamefont{{Cohen} et~al.}(2010)}]{cohen_etal10}
\bibinfo{author}{\bibfnamefont{T.}~\bibnamefont{{Cohen}}} \bibnamefont{et~al.},
  \bibinfo{journal}{\prd} \textbf{\bibinfo{volume}{82}},
  \bibinfo{pages}{056001} (\bibinfo{year}{2010}).

\bibitem[{\citenamefont{{Davoudiasl} et~al.}(2010)}]{davoudiasl_etal10}
\bibinfo{author}{\bibfnamefont{H.}~\bibnamefont{{Davoudiasl}}}
  \bibnamefont{et~al.}, \bibinfo{journal}{Phys. Rev. Lett.}
  \textbf{\bibinfo{volume}{105}}, \bibinfo{pages}{211304}
  (\bibinfo{year}{2010}).

\bibitem[{\citenamefont{{Falkowski} et~al.}(2011)\citenamefont{{Falkowski},
  {Ruderman}, and {Volansky}}}]{falkowski_etal11}
\bibinfo{author}{\bibfnamefont{A.}~\bibnamefont{{Falkowski}}},
  \bibinfo{author}{\bibfnamefont{J.~T.} \bibnamefont{{Ruderman}}},
  \bibnamefont{and}
  \bibinfo{author}{\bibfnamefont{T.}~\bibnamefont{{Volansky}}}
  (\bibinfo{year}{2011}), \eprint{[arXiv:1101.4936]}.

\bibitem[{\citenamefont{{Bernabei} et~al.}(2008)}]{dama08}
\bibinfo{author}{\bibfnamefont{R.}~\bibnamefont{{Bernabei}}}
  \bibnamefont{et~al.}, \bibinfo{journal}{Euro. Phys. J. C}
  \textbf{\bibinfo{volume}{56}}, \bibinfo{pages}{333} (\bibinfo{year}{2008}).

\bibitem[{\citenamefont{{Aalseth} et~al.}(2011{\natexlab{a}})}]{cogent11}
\bibinfo{author}{\bibfnamefont{C.~E.} \bibnamefont{{Aalseth}}}
  \bibnamefont{et~al.}, \bibinfo{journal}{Phys. Rev. Lett.}
  \textbf{\bibinfo{volume}{106}}, \bibinfo{pages}{131301}
  (\bibinfo{year}{2011}{\natexlab{a}}).

\bibitem[{\citenamefont{{Aalseth} et~al.}(2011{\natexlab{b}})}]{cogent11b}
\bibinfo{author}{\bibfnamefont{C.~E.} \bibnamefont{{Aalseth}}}
  \bibnamefont{et~al.}, \bibinfo{journal}{ArXiv e-prints}
  (\bibinfo{year}{2011}{\natexlab{b}}), \eprint{1106.0650}.

\bibitem[{\citenamefont{{Angloher} et~al.}(2011)\citenamefont{{Angloher},
  {Bauer}, {Bavykina}, {Bento}, {Bucci}, {Ciemniak}, {Deuter}, {von
  Feilitzsch}, {Hauff}, {Huff} et~al.}}]{cresst11}
\bibinfo{author}{\bibfnamefont{G.}~\bibnamefont{{Angloher}}},
  \bibinfo{author}{\bibfnamefont{M.}~\bibnamefont{{Bauer}}},
  \bibinfo{author}{\bibfnamefont{I.}~\bibnamefont{{Bavykina}}},
  \bibinfo{author}{\bibfnamefont{A.}~\bibnamefont{{Bento}}},
  \bibinfo{author}{\bibfnamefont{C.}~\bibnamefont{{Bucci}}},
  \bibinfo{author}{\bibfnamefont{C.}~\bibnamefont{{Ciemniak}}},
  \bibinfo{author}{\bibfnamefont{G.}~\bibnamefont{{Deuter}}},
  \bibinfo{author}{\bibfnamefont{F.}~\bibnamefont{{von Feilitzsch}}},
  \bibinfo{author}{\bibfnamefont{D.}~\bibnamefont{{Hauff}}},
  \bibinfo{author}{\bibfnamefont{P.}~\bibnamefont{{Huff}}},
  \bibnamefont{et~al.}, \bibinfo{journal}{[arXiv:1109.0702]}
  (\bibinfo{year}{2011}), \eprint{1109.0702}.

\bibitem[{\citenamefont{{Fitzpatrick} et~al.}(2010)\citenamefont{{Fitzpatrick},
  {Hooper}, and {Zurek}}}]{fitzpatrick_etal10}
\bibinfo{author}{\bibfnamefont{A.~L.} \bibnamefont{{Fitzpatrick}}},
  \bibinfo{author}{\bibfnamefont{D.}~\bibnamefont{{Hooper}}}, \bibnamefont{and}
  \bibinfo{author}{\bibfnamefont{K.~M.} \bibnamefont{{Zurek}}},
  \bibinfo{journal}{\prd} \textbf{\bibinfo{volume}{81}},
  \bibinfo{pages}{115005} (\bibinfo{year}{2010}), \eprint{1003.0014}.

\bibitem[{\citenamefont{{Chang} et~al.}(2010)}]{chang_etal10}
\bibinfo{author}{\bibfnamefont{S.}~\bibnamefont{{Chang}}} \bibnamefont{et~al.},
  \bibinfo{journal}{\jcap} \textbf{\bibinfo{volume}{8}}, \bibinfo{pages}{18}
  (\bibinfo{year}{2010}).

\bibitem[{\citenamefont{{Angle} et~al.}(2011)}]{xenon10_11}
\bibinfo{author}{\bibfnamefont{J.}~\bibnamefont{{Angle}}} \bibnamefont{et~al.}
  (\bibinfo{year}{2011}), \eprint{[arXiv:1104.3088]}.

\bibitem[{\citenamefont{{Spergel} and {Press}}(1985)}]{spergel_press85}
\bibinfo{author}{\bibfnamefont{D.~N.} \bibnamefont{{Spergel}}}
  \bibnamefont{and} \bibinfo{author}{\bibfnamefont{W.~H.}
  \bibnamefont{{Press}}}, \bibinfo{journal}{\apj}
  \textbf{\bibinfo{volume}{294}}, \bibinfo{pages}{663} (\bibinfo{year}{1985}).

\bibitem[{\citenamefont{{Gould}}(1992)}]{gould92}
\bibinfo{author}{\bibfnamefont{A.}~\bibnamefont{{Gould}}},
  \bibinfo{journal}{\apj} \textbf{\bibinfo{volume}{388}}, \bibinfo{pages}{338}
  (\bibinfo{year}{1992}).

\bibitem[{\citenamefont{{Cumberbatch} et~al.}(2010)}]{cumberbatch_etal10}
\bibinfo{author}{\bibfnamefont{D.~T.} \bibnamefont{{Cumberbatch}}}
  \bibnamefont{et~al.}, \bibinfo{journal}{\prd} \textbf{\bibinfo{volume}{82}},
  \bibinfo{pages}{103503} (\bibinfo{year}{2010}), \eprint{1005.5102}.

\bibitem[{\citenamefont{{McDermott} et~al.}(2011)\citenamefont{{McDermott},
  {Yu}, and {Zurek}}}]{mcdermott_etal11}
\bibinfo{author}{\bibfnamefont{S.~D.} \bibnamefont{{McDermott}}},
  \bibinfo{author}{\bibfnamefont{H.-B.} \bibnamefont{{Yu}}}, \bibnamefont{and}
  \bibinfo{author}{\bibfnamefont{K.~M.} \bibnamefont{{Zurek}}},
  \bibinfo{journal}{[arXiv:1103.5472]}  (\bibinfo{year}{2011}).

\bibitem[{\citenamefont{{Zentner}}(2009)}]{zentner09}
\bibinfo{author}{\bibfnamefont{A.~R.} \bibnamefont{{Zentner}}},
  \bibinfo{journal}{\prd} \textbf{\bibinfo{volume}{80}},
  \bibinfo{pages}{063501} (\bibinfo{year}{2009}).

\bibitem[{\citenamefont{{Press} and {Spergel}}(1985)}]{press_spergel85}
\bibinfo{author}{\bibfnamefont{W.~H.} \bibnamefont{{Press}}} \bibnamefont{and}
  \bibinfo{author}{\bibfnamefont{D.~N.} \bibnamefont{{Spergel}}},
  \bibinfo{journal}{\apj} \textbf{\bibinfo{volume}{296}}, \bibinfo{pages}{679}
  (\bibinfo{year}{1985}).

\bibitem[{\citenamefont{{Kippenhahn} and {Weigert}}(1994)}]{kippenhahn_weigert}
\bibinfo{author}{\bibfnamefont{R.}~\bibnamefont{{Kippenhahn}}}
  \bibnamefont{and}
  \bibinfo{author}{\bibfnamefont{A.}~\bibnamefont{{Weigert}}},
  \emph{\bibinfo{title}{{Stellar Structure and Evolution}}}
  (\bibinfo{year}{1994}).

\bibitem[{\citenamefont{{Paxton} et~al.}(2011)\citenamefont{{Paxton},
  {Bildsten}, {Dotter}, {Herwig}, {Lesaffre}, and {Timmes}}}]{mesa_instrument}
\bibinfo{author}{\bibfnamefont{B.}~\bibnamefont{{Paxton}}},
  \bibinfo{author}{\bibfnamefont{L.}~\bibnamefont{{Bildsten}}},
  \bibinfo{author}{\bibfnamefont{A.}~\bibnamefont{{Dotter}}},
  \bibinfo{author}{\bibfnamefont{F.}~\bibnamefont{{Herwig}}},
  \bibinfo{author}{\bibfnamefont{P.}~\bibnamefont{{Lesaffre}}},
  \bibnamefont{and} \bibinfo{author}{\bibfnamefont{F.}~\bibnamefont{{Timmes}}},
  \bibinfo{journal}{\apjs} \textbf{\bibinfo{volume}{192}}, \bibinfo{pages}{3}
  (\bibinfo{year}{2011}).

\bibitem[{\citenamefont{{Gould} and {Raffelt}}(1990)}]{gould_raffelt90b}
\bibinfo{author}{\bibfnamefont{A.}~\bibnamefont{{Gould}}} \bibnamefont{and}
  \bibinfo{author}{\bibfnamefont{G.}~\bibnamefont{{Raffelt}}},
  \bibinfo{journal}{\apj} \textbf{\bibinfo{volume}{352}}, \bibinfo{pages}{669}
  (\bibinfo{year}{1990}).

\bibitem[{\citenamefont{{Archambault} et~al.}(2009)}]{picasso09}
\bibinfo{author}{\bibfnamefont{S.}~\bibnamefont{{Archambault}}}
  \bibnamefont{et~al.}, \bibinfo{journal}{Phys. Lett. B}
  \textbf{\bibinfo{volume}{682}}, \bibinfo{pages}{185} (\bibinfo{year}{2009}).

\bibitem[{\citenamefont{{Gould}}(1987)}]{gould87a}
\bibinfo{author}{\bibfnamefont{A.}~\bibnamefont{{Gould}}},
  \bibinfo{journal}{\apj} \textbf{\bibinfo{volume}{321}}, \bibinfo{pages}{560}
  (\bibinfo{year}{1987}).

\bibitem[{\citenamefont{{Kumar}}(1963)}]{kumar63}
\bibinfo{author}{\bibfnamefont{S.~S.} \bibnamefont{{Kumar}}},
  \bibinfo{journal}{\apj} \textbf{\bibinfo{volume}{137}}, \bibinfo{pages}{1121}
  (\bibinfo{year}{1963}).

\bibitem[{\citenamefont{{Chabrier} and {Baraffe}}(2000)}]{chabrier_baraffe00}
\bibinfo{author}{\bibfnamefont{G.}~\bibnamefont{{Chabrier}}} \bibnamefont{and}
  \bibinfo{author}{\bibfnamefont{I.}~\bibnamefont{{Baraffe}}},
  \bibinfo{journal}{\araa} \textbf{\bibinfo{volume}{38}}, \bibinfo{pages}{337}
  (\bibinfo{year}{2000}).

\bibitem[{\citenamefont{{Kaiser} et~al.}(2002)}]{panstarrs02}
\bibinfo{author}{\bibfnamefont{N.}~\bibnamefont{{Kaiser}}}
  \bibnamefont{et~al.}, in \emph{\bibinfo{booktitle}{Society of Photo-Optical
  Instrumentation Engineers (SPIE) Conference Series}}, edited by
  \bibinfo{editor}{\bibnamefont{{J.~A.~Tyson \& S.~Wolff}}}
  (\bibinfo{year}{2002}), vol. \bibinfo{volume}{4836} of
  \emph{\bibinfo{series}{Presented at the Society of Photo-Optical
  Instrumentation Engineers (SPIE) Conference}}, pp. \bibinfo{pages}{154--164}.

\bibitem[{\citenamefont{{LSST Science Collaborations}}(2009)}]{lsstbook09}
\bibinfo{author}{\bibnamefont{{LSST Science Collaborations}}}
  (\bibinfo{year}{2009}), \eprint{[arXiv:0912.0201]}.

\bibitem[{\citenamefont{{Refregier} et~al.}(2010)}]{euclid10}
\bibinfo{author}{\bibfnamefont{A.}~\bibnamefont{{Refregier}}}
  \bibnamefont{et~al.} (\bibinfo{year}{2010}), \eprint{[arXiv:1001.0061]}.

\bibitem[{\citenamefont{{Marley} and {Leggett}}(2009)}]{jwstbd09}
\bibinfo{author}{\bibfnamefont{M.~S.} \bibnamefont{{Marley}}} \bibnamefont{and}
  \bibinfo{author}{\bibfnamefont{S.~K.} \bibnamefont{{Leggett}}},
  \emph{\bibinfo{title}{{The Future of Ultracool Dwarf Science with JWST}}}
  (\bibinfo{year}{2009}), pp. \bibinfo{pages}{101--+}.

\bibitem[{\citenamefont{{Bruch} et~al.}(2009)}]{bruch_etal09}
\bibinfo{author}{\bibfnamefont{T.}~\bibnamefont{{Bruch}}} \bibnamefont{et~al.},
  \bibinfo{journal}{Phys. Lett. B} \textbf{\bibinfo{volume}{674}},
  \bibinfo{pages}{250} (\bibinfo{year}{2009}).

\bibitem[{\citenamefont{{Purcell} et~al.}(2009)\citenamefont{{Purcell},
  {Bullock}, and {Kaplinghat}}}]{purcell_etal09}
\bibinfo{author}{\bibfnamefont{C.~W.} \bibnamefont{{Purcell}}},
  \bibinfo{author}{\bibfnamefont{J.~S.} \bibnamefont{{Bullock}}},
  \bibnamefont{and}
  \bibinfo{author}{\bibfnamefont{M.}~\bibnamefont{{Kaplinghat}}},
  \bibinfo{journal}{\apj} \textbf{\bibinfo{volume}{703}}, \bibinfo{pages}{2275}
  (\bibinfo{year}{2009}).

\bibitem[{\citenamefont{{Meylan} and {Heggie}}(1997)}]{meylan_heggie97}
\bibinfo{author}{\bibfnamefont{G.}~\bibnamefont{{Meylan}}} \bibnamefont{and}
  \bibinfo{author}{\bibfnamefont{D.~C.} \bibnamefont{{Heggie}}},
  \bibinfo{journal}{\aapr} \textbf{\bibinfo{volume}{8}}, \bibinfo{pages}{1}
  (\bibinfo{year}{1997}).

\bibitem[{\citenamefont{{Conroy} et~al.}(2010)\citenamefont{{Conroy}, {Loeb},
  and {Spergel}}}]{conroy_etal10}
\bibinfo{author}{\bibfnamefont{C.}~\bibnamefont{{Conroy}}},
  \bibinfo{author}{\bibfnamefont{A.}~\bibnamefont{{Loeb}}}, \bibnamefont{and}
  \bibinfo{author}{\bibfnamefont{D.}~\bibnamefont{{Spergel}}}
  (\bibinfo{year}{2010}), \eprint{[arXiv:1010.5783]}.

\bibitem[{\citenamefont{{Simon} et~al.}(2011)}]{simon_etal11}
\bibinfo{author}{\bibfnamefont{J.~D.} \bibnamefont{{Simon}}}
  \bibnamefont{et~al.}, \bibinfo{journal}{\apj} \textbf{\bibinfo{volume}{733}},
  \bibinfo{pages}{46} (\bibinfo{year}{2011}).

\bibitem[{\citenamefont{{van Dokkum} and {Conroy}}(2010)}]{vandokkum_conroy10}
\bibinfo{author}{\bibfnamefont{P.~G.} \bibnamefont{{van Dokkum}}}
  \bibnamefont{and} \bibinfo{author}{\bibfnamefont{C.}~\bibnamefont{{Conroy}}},
  \bibinfo{journal}{\nat} \textbf{\bibinfo{volume}{468}}, \bibinfo{pages}{940}
  (\bibinfo{year}{2010}).

\end{thebibliography}

\end{document}